\documentclass[conference]{IEEEtran}

\usepackage[english]{babel}
\usepackage{amssymb,textcomp}
\usepackage[usenames]{color}
\usepackage{listings}
\usepackage{paralist}
\usepackage{url}
\usepackage{cite}
\usepackage{xspace}
\usepackage{graphicx}
\usepackage{multirow}
\usepackage{makecell}
\usepackage{fourier} 
\usepackage{array}
\usepackage{array,booktabs}
\usepackage[pdftex,colorlinks=true,pdfstartview=FitV,
 linkcolor=black,citecolor=black,urlcolor=black,bookmarks=false]{hyperref}
\usepackage{needspace}

\usepackage{tikz}
 
\usepackage{ifthen}
\usepackage[normalem]{ulem} 
\usepackage{xcolor}

\newboolean{showedits}
\setboolean{showedits}{true} 
\ifthenelse{\boolean{showedits}}
{
	\newcommand{\del}[1]{\textcolor{red}{\sout{#1}}} 
}{
	\newcommand{\del}[1]{} 
	
}
\usepackage{amssymb}
\newboolean{showcomments}
\setboolean{showcomments}{true}
\newcommand{\id}[1]{$-$Id: scg-llncs.tex 30911 2010-02-05 10:21:47Z oscar $-$}

\ifthenelse{\boolean{showcomments}}
{\newcommand{\nbc}[3]{
 {\colorbox{#3}{\bfseries\sffamily\scriptsize\textcolor{white}{#1}}}
 {\textcolor{#3}{\sf\small$\blacktriangleright$\textit{#2}$\blacktriangleleft$}}}
 }
{\newcommand{\nbc}[3]{}
 \renewcommand{\del}[1]{} 
 }

\newcommand{\ie}{\emph{i.e.},\xspace}
\newcommand{\eg}{\emph{e.g.},\xspace}

\newboolean{isblinded}
\setboolean{isblinded}{false}
\ifthenelse{\boolean{isblinded}}
{\newcommand\blind[1]{BLINDED\xspace}}
{\newcommand\blind[1]{#1\xspace}}

\newcounter{col}
\newcommand\col{\stepcounter{col}\emph{(COL\thecol{})}}
\newcounter{com}
\newcommand\com{\stepcounter{com}\emph{(COM\thecom{})}}
\newcounter{mob}
\newcommand\mob{\stepcounter{mob}\emph{(MOB\themob{})}}
\newcounter{med}
\newcommand\med{\stepcounter{med}\emph{(MED\themed{})}}
\newcounter{pri}
\newcommand\pri{\stepcounter{pri}\emph{(PRI\thepri{})}}
\newcounter{emb}
\newcommand\emb{\stepcounter{emb}\emph{(EMB\theemb{})}}
\newcounter{per}
\newcommand\per{\stepcounter{per}\emph{(PER\theper{})}}
\newcounter{mul}
\newcommand\mul{\stepcounter{mul}\emph{(MUL\themul{})}}

\usepackage{wasysym}
\usepackage{cite}
\usepackage{nth}
\usepackage{scrextend}
\usepackage[ancient,keeplastbox]{flushend}
\usepackage{wrapfig}
\usepackage{subcaption}
\usepackage{colortbl}
\usepackage[all]{nowidow}

\begin{document}
\font\myfont=cmr12 at 24pt
\title{{\myfont Unleashing the Potentials of Immersive Augmented Reality for Software Engineering}}

\author{\IEEEauthorblockN{\blind{Leonel Merino}\IEEEauthorrefmark{1},
\blind{Mircea Lungu}\IEEEauthorrefmark{2},
\blind{Christoph Seidl}\IEEEauthorrefmark{2}}
\IEEEauthorblockA{\IEEEauthorrefmark{1}\blind{VISUS, University of Stuttgart, Germany}}
\IEEEauthorblockA{\IEEEauthorrefmark{2}\blind{Computer Science Department, IT University of Copenhagen, Denmark}}}

\maketitle

\begin{abstract}

In immersive augmented reality (IAR), users can wear a head-mounted display to see computer-generated images superimposed to their view of the world. IAR was shown to be beneficial across several domains, \eg automotive, medicine, gaming and engineering, with positive impacts on, \eg collaboration and communication. We think that IAR bears a great potential for software engineering but, as of yet, this research area has been 
neglected.
In this vision paper, we elicit potentials and obstacles for the use of IAR in software engineering. We identify possible areas that can be supported with IAR technology by relating commonly discussed IAR improvements to typical software engineering tasks. We further demonstrate how innovative use of IAR technology may fundamentally improve typical activities of a software engineer through a comprehensive series of usage scenarios outlining practical application. Finally, we reflect on current limitations of IAR technology based on our scenarios and sketch research activities necessary to make our vision a reality. We consider this paper to be relevant to academia and industry alike in guiding the steps to innovative research and applications for IAR in software engineering.
\end{abstract}

\section{Introduction}
In immersive augmented reality (IAR), a user's view of the world is altered by superimposing computer-generated images. Usually, IAR is realized by means of a head-mounted display (HMD) and natural user interfaces (NUI). That is, users can interact with the immersive environment using  hand gestures as well as head and body movements. Indeed, the use of IAR was shown to be effective across various research fields. To name a few examples, in automotive, researchers  used IAR to enable drivers to use visualizations projected on a car's windshield to analyze auxiliary information while keeping their attention on the road~\cite{6671779}. In medicine, researchers found that IAR allows doctors to make more effective decisions based on computer-generated images that represent organs of a patient during surgery~\cite{6671780}. Consequently, we ask whether the benefits of employing IAR found in other fields can be transferred to tasks involved in software engineering.     
\vspace{0.5em}

\noindent\textbf{Related work}. Although little research has focused on investigating the potentials of using IAR to support software engineering, we consider these initial works to be important contributions: For instance, IAR has been employed to support the analysis of software architectures based on visualization~\cite{schreiber2019visualization} and conversational interfaces~\cite{seipel2019adopting}. IAR has also been used to navigate 3D visualizations to support software comprehension~\cite{Meri18c} and software performance~\cite{Meri19a} tasks. A study analyzed using a Microsoft HoloLens device to explore UML diagrams~\cite{mikkelsen2017exploring}. Another study found that IAR can boost motivation of students in software engineering education~\cite{reuter2019using}. Yet another study proposed to help create a project workspace supported by IAR~\cite{sharma2019extended}. 
In a broader context, some studies~\cite{Meri17b,romano2019city,rodrigues2016visar3d} found that software visualizations displayed in immersive virtual reality promote users' engagement. Other technologies such as multi-touch have been used to augment code reviews~\cite{Anslow-multitouch,1753548}. 
In summary, we think that the impact of a positive user experience of immersive technologies, in particular IAR, on user performance warrants further examination~\cite{Shar18a}.  
\begin{figure}[t]
	\centering
	\includegraphics[width=\linewidth]{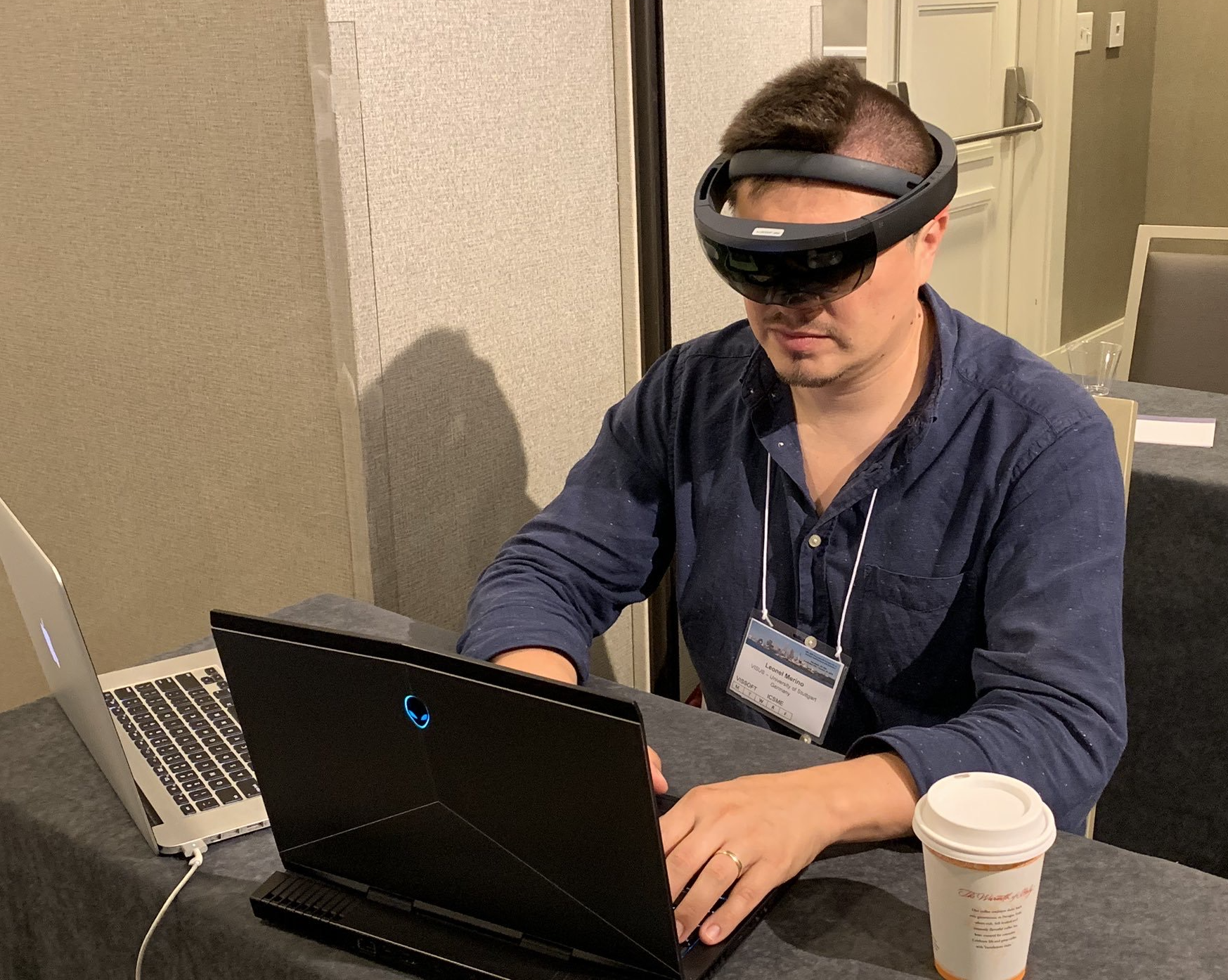}
	\caption{A software engineer wears a Microsoft HoloLens to extend the space of his laptops' screens by superimposing computer-generated images in the physical environment.}
	\label{fig:teaser}
	\vspace{-1.0cm}
\end{figure} 

To examine what can be the role of IAR in the future of software engineering, we follow four steps:
\begin{inparaenum}[\itshape (i)\upshape] 
    \item we identify a set of dimensions that are frequently found amongst the variables examined in evaluations of IAR approaches in other fields, 
    \item we analyze how these dimensions could impact software engineering tasks,   
    \item we envision usage scenarios that exemplify the use of IAR amongst the processes of software development, and
    \item we discuss benefits and limitations that we foresee will be associated with the adoption of IAR in software engineering. 
\end{inparaenum}



In our envisioned future, software engineers wear IAR headsets as shown in Figure~\ref{fig:teaser} (\eg Microsoft HoloLens\footnote{\url{https://www.microsoft.com/en-us/hololens}}) --an optical see-through headset that corresponds to a fully capable modern computer. That is, the device has processing power, persistent storage, network connectivity, and I/O features such as a display, camera, and microphone. In this vision, there is a continuum of degrees at which IAR could be integrated into the software engineering. The most lightweight integration we envision is using IAR as a mere companion output device that extends the graphical user interface space offered by the computer screen and keeping unchanged the previous display medium and interaction methods (\ie computer screen, keyboard and mouse).
On the opposite side, we envision adopting IAR as the main input/output device. That means to completely replace the computer screen and mouse by exploiting the technology included in the IAR device. For instance, software engineers could use cameras in the device to identify elements in the physical reality (\eg text on a whiteboard, icons on a screen, colleagues' faces) that can be augmented with digital representations (\eg lightweight visualizations, infographics), and sensors that can be used to track hand gestures as well as head and eye movements to support interactions. 



\section{A Preliminary Framework}
\label{sec:framework}
We are convinced that there is a great potential of IAR for novel applications in software engineering that is yet untapped. We analyze the potential benefits through the lens of a framework based on aspects mentioned frequently in research literature on augmented reality~\cite{kim2018revisiting} and visualization ~\cite{isenberg2013systematic}. 
Even though these aspects do not focus core software development artifacts directly, we think that software engineering comprises many activities that are inherently based on human/machine or human/human interaction.
Hence, we conjecture that all these aspects have the potential to positively impact specific software engineering activities. Even though we present a comprehensive list, we do not claim completeness for the identified aspects. Instead, we aim at setting a common and relevant ground that can be used to reflect on how IAR can support software engineering. 
\vspace{0.2em}


\noindent\textbf{Collaboration.} 
Various studies have investigated the use of augmented reality to facilitate collaboration in teams to address tasks in multiple scenarios, such as investigating the effects of providing real-time heart rate feedback to remote collaborators for improving their user experience ~\cite{8613762} or discussing how to improve sharing and watching live 360 panorama based remote collaborative experiences through visualizations in IAR ~\cite{8613761}. 
In software engineering, for instance, IAR could be used to help co-located collaborators to navigate room-scaled 3D software visualizations.   
\vspace{0.2em}

\noindent\textbf{Communication.} 
Other studies analyzed the communicative value of an augmented reality technique to goals such as teaching or presentation. E.g. the use of augmented visual communication cues to improve users' experience when collaborating in a shared remote task space ~\cite{6948412} 
or 
using eye tracking to examine how avatar-mediated communication in shared virtual environments to promote social presence and emotions ~\cite{1753481}.
In software engineering, IAR could, for instance, support an operations engineer who requires orientation to find a server in a data center.   
\vspace{0.2em}

\noindent\textbf{Embodiment.} 
The impact of the experience that comes from the sensory-motor capacities of the body into human cognition is also frequently present across several studies in augmented reality research. For example, a study~\cite{8613756} examined how visual embodiment and social behavior influence perception of intelligent virtual agents using augmented reality. Another study~\cite{1978951} analyzed users' behavior when interacting with each other through a remote embodied space based on a high fidelity immersive environment that presents full scale and real-time images of co-participants.
In software engineering, embodiment promoted by IAR could help to establish a more dynamic working environment and overcome sedentary behavior.   
\vspace{0.2em}

\noindent\textbf{Mediated reality.} 
A frequent topic of interest in augmented reality research is mediated reality, which relates to changing the appearance of physical objects and scenes. For instance, a study~\cite{bach2017hologram} investigated the effectiveness of interactive exploration of 3D visualizations in immersive tangible augmented reality. Another study~\cite{3025795} presented an approach in augmented reality that alters the perception of the appearance of a physical object by controlling visual aspects of its surrounding space. 
Software engineers could use IAR with objects of daily use by adding to them digital enhancements that blend with reality.
\vspace{0.2em}

\noindent\textbf{Mobility.} 
The mobile capabilities of augmented reality approaches also are a frequent subject of study. For example, a study~\cite{8458443} presented an approach that enables a dynamic 3D reconstruction of indoor and outdoor scenes of everyday environments based on cameras worn by a user in IAR. Another study~\cite{2208677} analyzed awareness of surroundings, empowerment, positive surprise, amazement and fascination, immersion, and social connectivity of early stage mobile augmented reality applications.
In software engineering, for instance, IAR could boost the productiveness of developers waiting in an airport, by enabling them to see multiple monitors as if they were at their office.
\vspace{0.2em}

\noindent\textbf{Multi-device.}
A frequent topic is also the study of how we can interact with multiple devices of various characteristics. Some devices can have a screen (of various sizes), whereas some others project images on surfaces or involve see-through displays. For instance, a study~\cite{2208297} presented a metaphor for multi-device interaction with handhelds and large displays based on position, size, and orientation control of a handheld optical projector. Another study~\cite{8613758} investigated the use of IAR to extend the available screen of smartphones beyond their physical limits with a co-planar virtual surface.
Software engineers could use IAR to integrate such interfaces displayed in multiple devices.
\vspace{0.2em}

\noindent\textbf{Pervasiveness. }
In IAR, users can have a continuous experience based on mobility and multi-device interaction. In consequence, some studies investigated the impact of such pervasiveness. For instance, a study~\cite{7435333} discussed pervasive augmented reality of scenarios that provide a continuous and multi-purpose user experience by adapting the system based on the changing requirements and constraints of users' current context. Another study~\cite{Meri19a} presented a pervasive visualization approach that uses IAR to enhance software performance awareness. 
Indeed, software engineers could benefit, for example, from omnipresent views of multiple information, \eg status of a continuous integration system, communication within a team.
\vspace{0.2em}

\noindent\textbf{Privacy.}
An important topic of interest is how the technologies involved in augmented reality can be used to promote privacy protection. For instance, a study~\cite{1753548} presented a virtual elastic "tether" for mobile devices that allows groups to have quick, simple, and privacy-preserving pedestrian meetups. Another study~\cite{yagi2017diminished} presented a method that hides people from image sequences taken with a hand-held camera for privacy protection. In software engineering, IAR could be beneficial for projects that involve confidential information by preventing access to unauthorized viewers.  

We observe that the framework defined by these eight aspects can be used to identify research gaps by analyzing the applicability of IAR to particular software engineering tasks. We elaborate on such analysis by means of visionary usage scenarios.  




\section{Usage Scenarios}


We now present usage scenarios for an envisioned future life of software engineers using IAR. Each scenario highlights specific examples that stress the support of IAR. Even though state-of-the-art software development processes are diverse, they contain common fundamental activities, \eg requirements engineering, software architecture or software design, implementation. We structure our usage scenarios along those activities highlighting improvements through IAR for each one of them. In Section~\ref{sec:discussion}, we relate aspects of our conceptual framework from Section~\ref{sec:framework} to each one of these scenarios, for which we introduce markers in this section, \eg \textit{(MOB1)} for the first mobility example.

\vspace{0.2em}

\noindent\textbf{Requirements Engineering.}
Grace is a requirements engineer. Her work consists of understanding business needs to produce a specification of the requirements of a software system. She wears an IAR headset that supports her on various requirements engineering tasks. 
One such task is \emph{augmented requirements elicitation.} While interviewing a stakeholder~\mob~
\begin{inparaenum}[\itshape (i)\upshape] 
    \item she can obtain an interactive view of the domain model, which reacts by highlighting nodes as the customer is talking~\per, and
    \item she is able to glimpse at the domain model concepts, which helps her formulate more comprehensive questions~\col. 
\end{inparaenum}
Another task is \emph{augmented requirements analysis.} 
Grace likes taking notes on paper and drawing diagrams using a whiteboard in her office. The camera of her IAR headset allows her to digitize hand written notes to boost their analysis~\med. Using IAR, she can categorize and organize notes by spatially rearranging them as well as decomposing notes into more atomic ones~\emb. Then, notes can be linked to produce a view that supports Grace to trace requirements to their source. 

\vspace{0.2em}

\noindent\textbf{Software Design.} 
Pete, a software designer, wears his IAR headset to virtually project a UML class diagram of a software system on a wall at his office to use the ample surface for a diagram overview~\med. He needs to identify what classes can be reused and how the system should be extended. He stands in front of the wall~\emb~and adds classes to the diagram by drawing boxes and links from the boxes to their corresponding hierarchies with a marker. The marker does not contain ink but virtually traces movements and gives haptic feedback~\med, which makes his drawing more precise. He also uses the microphone in the IAR device and speech recognition to enter the name of the newly added classes. Once finished, he uses a hand gesture swiping the diagram onto his computer screen to create a source code skeleton of the new classes in the IDE~\mul. 

\vspace{0.2em}

\noindent\textbf{Implementation.} 
Alice, a software developer, is ready to start the implementation. On her desk, there is only a keyboard. Although the keyboard exists in physical reality, labels on the keys are displayed in augmented reality~\med. 
The IAR headset functions as \emph{augmented navigation}~\med~that enables her to efficiently navigate through the code while helping her create a mental map of the source code~\cite{wettel-habitability}~\per.
To the left of her keyboard she projects the most recently visited files as well as most likely files to which she might want to navigate based on a degree-of-interest model~\cite{Kersten-Mylar}.
To the right of her keyboard, she keeps a series of visualizations of the source code visible only through her IAR headset. 
Interacting with these virtual navigation affordances is faster and more natural than using the mouse to "peck" tabs from history and allows her to build a spatial representation of the source code. 
On the wall behind her screen, she displays context-aware documentation that is related to the code at hand~\mob. This is where she sometimes brings up the traceability matrix for the project, or the project backlog.
During her long commute by train Alice is using her IAR as an the main output device: besides the fact that she can continue her work seamlessly from the office, she also does not have to worry about the privacy of her data being compromised by passers by looking at her screen ~\pri. 

\vspace{0.2em}

\begin{table*}[ht!]
\caption{A summary of aspects of IAR with potential benefits for software engineering.}
\label{tab:analysis}
\setlength\tabcolsep{4.5pt}
\setlength\extrarowheight{-5pt}
\begin{tabular}{lllllllll}\toprule
 & Collaboration & Communication & Embodiment & Mediated Reality & Mobility & Multi-device & Pervasiveness & Privacy \\\midrule
Requirements Engineering    & COL1 & -- & EMB1 & MED1 & MOB1  & -- & PER1 & -- \\
Software Design             & -- & -- & EMB2 & MED2--3 & -- & MUL1 & -- & -- \\
Implementation     & -- & -- & -- & MED4--5 & MOB2 & -- & PER2 & PRI1\\
DevOps                      & --  & COM1 & --- & MED6--8 & MOB3--4 & -- &  -- & --\\
Software Testing            & COL2  & -- & EMB3--4 & MED9--10 & -- & -- & PER3  & --\\
Software Maintenance        & COL3  & -- & EMB5 & MED11--12 & MOB5 & -- & -- & --\\
\bottomrule
\vspace{-3em}
\end{tabular}
\end{table*}

\noindent\textbf{DevOps.}
Don is a DevOps engineer working on a desktop setup with mouse and keyboard but augmented with an IAR headset. On the desk, he keeps a set of \emph{augmented commands} --a series of extra programmable input areas represented as virtual buttons visible only through his IAR headset~\med. They capture repetitive actions he needs frequently, such as building and redeploying an application on the test server\footnote{This is inspired by the original setup that Douglas Engelbart envisioned for his {\em Mother of All Demos} presentation but it is not limited to only five keys and the user does not have to memorize the meaning of the keys.\\\url{https://www.youtube.com/watch?v=yJDv-zdhzMY}}.
A message arrives on his IAR headset~\com~and Don is warned that the application deployed in production is having performance issues. The components of the application are deployed in multiple servers in the data center. An alarm warns Don that one of the servers is shut down, and he needs to configure the load balancer on site~\mob.  
Don goes to the data center~\mob~and uses his IAR headset to retrieve a visualization of the traffic overlaid on the physical servers. He filters the traffic of applications of other customers to isolate the traffic between the components of the application of his client. In the visualization, a glyph rendered on top of physical racks shows information of deployed components of the application in a server~\med. Colored thick edges placed on the floor represent metrics of the traffic in the network amongst servers (\eg bandwidth, latency)~\med. Don spots the cause of the issue. He uses a hand gesture to deploy an instance of the involved component in a less demanded server and reroute network traffic.
\vspace{0.2em}

\noindent\textbf{Software Testing.}
Craig, is a quality assurance engineer. He keeps a view of the debugger window on the wall on his right. With a slight head movement, he can move from writing tests to running them~\emb. To obtain a pervasive view of the health of the collaborative project~\col, he uses a city visualization that he places virtually on his desk close to the keyboard. In the city, each building represents a test class. Buildings close to each other in a district represent test suites. Empty districts represent source code still not covered by tests. In the dynamic visualization (that is connected to the continuous integration server), buildings become green whenever a test passed and red otherwise~\per.
Craig is writing new tests. He likes to use the whiteboard. The IAR device is able to capture the written code. When he finishes writing a new test and the code does not compile, his hand written code is highlighted (in IAR) to point to a syntax error~\med. IAR also supports Craig by displaying a list of keywords (\eg existing interface methods) simulating an auto-complete feature. When the code finally compiles, a red button appears (in IAR) on the whiteboard next the the code~\med. He now can push the button to run the test. He feels good of adopting this embodied approach to write tests~\emb.
\vspace{0.2em}

\noindent\textbf{Software Maintenance.}
Marcus and Bob are software maintainers discussing a strategy to package the software system for distribution~\col. They stand at the whiteboard to write snippets of a build configuration. Working at the whiteboard makes them feel more creative and open to engage in a conversation~\emb. They wear IAR headsets and share a collaborative application that uses their headsets' front camera to capture and parse the configuration description. The application can display in augmented reality extended information related to the configuration that Marcus is writing~\med, \eg transitive library versions. The application also informs them in real time when the package name they write is sufficiently unique, so they can save time by not having to write it fully on the whiteboard.
The application also enables Marcus to compile the system using his configuration. Suddenly, Marcus and Bob want to relocate to the cafeteria. As the IAR application captured the specified configuration, they are able to continue their discussion anywhere~\mob. For instance, they can use a wall in the cafeteria or they can even display the specification floating in the air~\med. They could choose between using a marker and writing on a glass wall in the cafeteria to completely immerse in augmented reality.
\section{Discussion}
\label{sec:discussion}
A summary of the aspects of IAR that could benefit software engineering is presented in Table~\ref{tab:analysis}. We observe that embodiment and mediated reality crosscut almost all software engineering processes. That is, IAR could boost cognitive abilities of software engineers by involving natural user interfaces such as hand gestures and head movements. Software engineers could also benefit from manipulating visual perception by involving other senses, \eg haptic. 

We describe usage scenarios that exemplify multiple benefits of adopting IAR in software engineering. However, we observe that IAR adoption implies several challenges as well.  
  There are challenges for researchers who will need to investigate new domain-specific interaction techniques for the domains of software modeling and code writing. Although some of the applications we have envisioned in this paper are specific to working with code (\eg augmented auto-complete on the whiteboard), we think applications for IAR in software engineering are not limited to it. We reflect that an implication of using IAR in software engineering will be investigating to which extent immersion brings benefits to particular tasks, and in which cases immersion could represent a drawback.
  There are also challenges for software engineers, \eg to modify their behavior and adopt natural user interfaces such as head and eye movements to interact with IAR interfaces. 
  Finally, there are challenges for society that will have to deal with software engineers wearing IAR devices in public spaces. An implication of it could be requiring to define protocols for using IAR devices in public spaces, \eg to warn when they are recording. 
%
\vspace{-0.5em}

\section{Conclusion}
We describe our vision of how immersive augmented reality (IAR) can support tasks involved in software engineering. To this end, we define a preliminary framework of eight relevant aspects to software engineering, which are frequently found in evaluations of IAR research. We elaborate on envisioned usage scenarios that span various processes involved in software engineering. Finally, we discuss the implications of using IAR on a daily basis. In the future, we plan to further our investigation and realize such blending of IAR into software engineering.

\section*{Acknowledgments}
\blind{Merino acknowledges funding by the Deutsche Forschungsgemeinschaft (DFG, German Research Foundation) -- Project-ID 251654672 -- TRR 161.}

 \bibliographystyle{IEEEtran}
 \bibliography{idear}
\end{document}